\documentstyle[12pt,twoside,fleqn,epsfig,espcrc1]{article}
\begin{document}

\title{The $s$-wave repulsion and deeply bound pionic atoms:
fact and fancy\thanks{Supported in part by the Israel Science Foundation.}}

\author{E. Friedman\address[RI]{Racah Institute of Physics, the Hebrew
        University, Jerusalem 91904 Israel} and A. Gal\addressmark[RI]}
\maketitle

\begin{abstract}
Fits to a large data set of pionic atoms show that the `missing'
$s$-wave repulsion is accounted for when a density dependence suggested
recently by Weise is included in the isovector term of the $s$-wave
pion optical potential. The importance of using large data sets is
demonstrated and the role of deeply bound pionic atom states is discussed.
\end{abstract}

\section{INTRODUCTION}

Conventional phenomenological analyses of strong interaction effects
in pionic atoms yield unexpectedly sizable $s$-wave repulsion \cite{BFG97}.
Weise suggested that partial restoration of chiral symmetry in dense matter
leads to enhancement of the free pion-nucleon isovector $s$-wave
amplitude $b_1$ which could explain the anomalous repulsion \cite{Wei01}.
The origin and magnitude of this enhancement have been discussed recently
\cite{KKW02,CEO02}. Weise's suggestion was tested by Friedman \cite{Fri02}
using a large set of pionic-atom data and was found to indeed account for
most of the anomaly. This analysis has been extended significantly by
increasing the number of data points from 60 to 106 and including several
series of isotopes. Various sets of nucleon densities were used in addition
to testing corrections due to a relativistic impulse approximation.
Here we briefly review the results of the extended analysis \cite{Fri02a}
and, furthermore, focus attention on the significance of fits to large data
sets and on the role played by the `deeply bound' pionic atom states,
particularly for the purpose of extracting the in-medium isovector $\pi N$
$s$-wave amplitude $b_1(\rho )$.

\section{RESULTS}
\subsection{Implementation of Weise's density-dependence for $b_1$}

\begin{figure}
\begin{minipage}[t]{76mm}
\epsfig{file=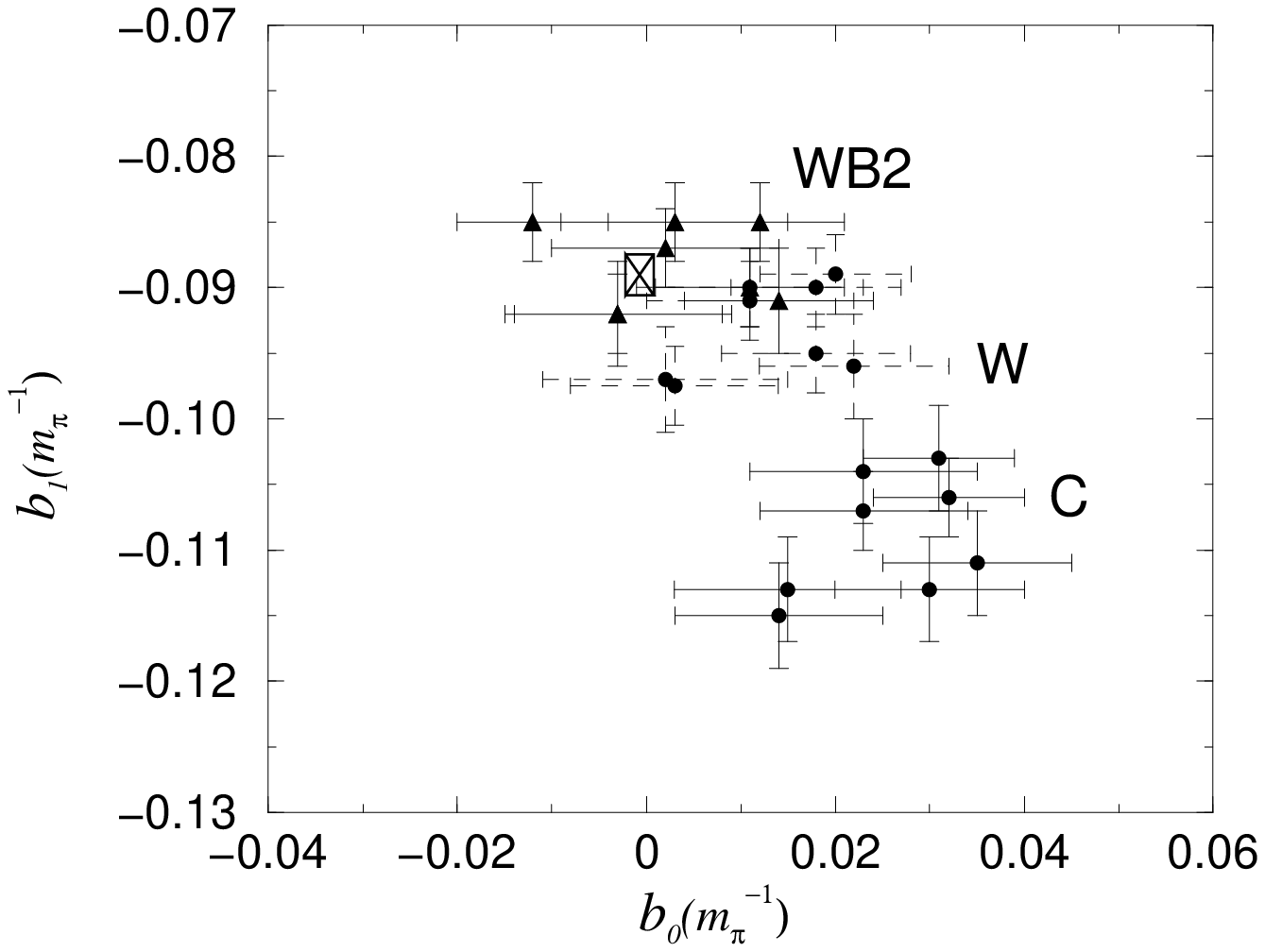,height=75mm,width=76mm}
\caption{Best-fit $b_0$ and $b_1$ for 106 data points \cite{Fri02a}.
X marks free-space values \cite{SBG01}.}
\label{fig:b0b1}
\end{minipage}
\hspace{\fill}
\begin{minipage}[t]{76mm}
\epsfig{file=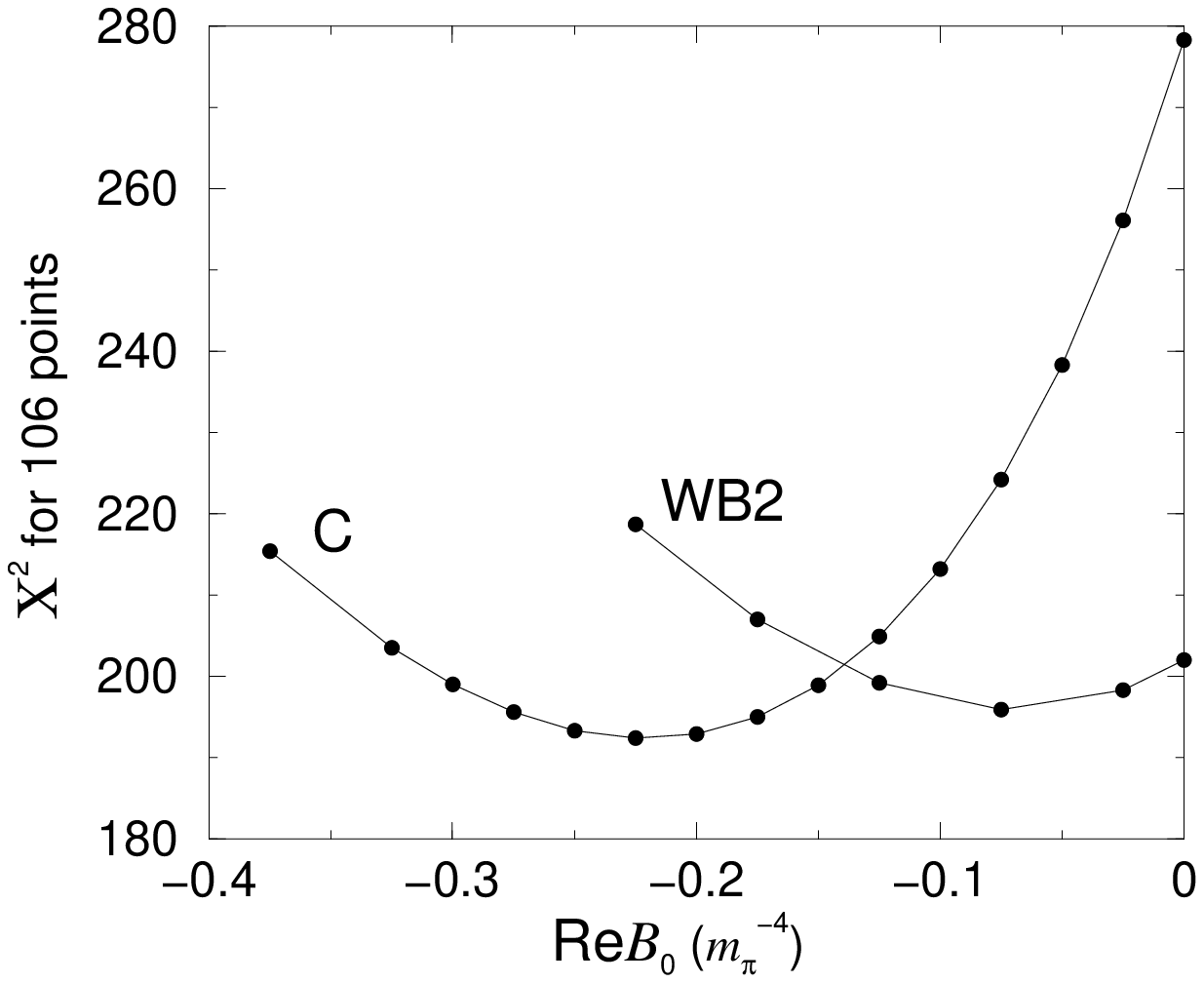,height=75mm,width=76mm}
\caption{Total $\chi ^2$ vs. Re$B_0$ for the C and WB2
comprehensive-data fits \cite{Fri02a}.}
\label{fig:para}
\end{minipage}
\end{figure}

Highlights of the results of Ref. \cite{Fri02a} are displayed in
Fig. \ref{fig:b0b1} for three potentials: (i) a conventional potential (C)
with $s$-wave component
\begin{eqnarray} \label{EE1s}
2{\mu}V^{(s)}_{\rm opt}(r) & = &
-4\pi(1+\frac{\mu}{M})\{\bar b_0[\rho_n(r)+\rho_p(r)]
  +b_1[\rho_n(r)-\rho_p(r)] \} \nonumber \\
& &  -4\pi(1+\frac{\mu}{2M})4B_0\rho_n(r) \rho_p(r);~~~~
\bar b_0 = b_0 - \frac{3}{2\pi}(b_0^2+2b_1^2)k_F(r),
\end{eqnarray}
where $k_F(r)$ is the local Fermi momentum. The second order term is included
in $\bar b_0$ because of the extremely small value of the isoscalar amplitude
$b_0$ \cite{SBG01}; (ii) the potential C where the isovector amplitude $b_1$
is made density dependent according to Ref. \cite{Wei01} (W, dashed);
and (iii) the latter potential where, in addition, a relativistic impulse
approximation density-dependence effect is included (WB2).
Eight points are given for each potential, corresponding to different
models for the nuclear densities, different values for the rms radii
of the neutron distributions and to different ways of handling the
$p$-wave part of the potential. All the fits are equally good, with
$\chi ^2 /F$ between 1.90 and 1.95. The figure shows clustering of points
with particularly well-defined values for the isovector amplitude $b_1$
(constant for C, and at $\rho = 0$ for W and WB2). 
It is clearly seen that the conventional
model C disagrees with the experimental free pion-nucleon values
\cite{SBG01} which are marked as a boxed X, particularly with regard
to $b_1$, whereas the Weise prescription W removes most of the discrepancy,
notably for $b_1$. A close agreement is observed for the WB2 model.
An important consequence of fitting to a large data base,
106 point in the present case, is that $b_1$ is determined extremely well
with as small errors as $\pm {0.004}~m_\pi ^{-1}$.
The errors on $b_0$ are larger, typically $\pm {0.010}~m_\pi ^{-1}$,
due to including a quadratic-density dispersive
term (with parameter Re$B_0$) in the potential \cite{Fri02a}.
The errors on $b_0$ could be {\it artificially} reduced by
excluding such a term, but only at the unbearable cost of increasing
$\chi ^2$ by 90 units for potential C, as is clearly seen in
Fig. \ref{fig:para}. This obviously means that such a dispersive term is
mandated by the data, contrary to recent claims \cite{KYa01,GGG02a}.
The combined effect of $b_0$, $b_1$ and Re$B_0$, irrespective of which
best-fit potential is adopted, is to produce a net repulsion of about
30 MeV inside of heavy nuclei such as Pb \cite{Fri02a}; this is essentially
the content of the phrase `anomalous repulsion' for pionic atoms. We note that
the size of the best-fit Re$B_0$, which is unacceptably large for C,
decreases gradually upon introducing the density dependence appropriate
to W and to WB2, becoming comparable for WB2 to the size of Im$B_0$
(0.054$\pm$0.002 $m_{\pi}^{-4}$) as appropriate to a dispersive term.

\subsection{Global vs. partial data-set analyses}

In order to assess the significance of using a large data base, we have
studied the consequences of using severely reduced data sets. Following
Yamazaki et al. \cite{KYa01,GGG02a},
we chose the 1$s$ states in the $N=Z$ nuclei
$^{12}$C, $^{14}$N, $^{16}$O, $^{20}$Ne, $^{24}$Mg and $^{28}$Si to which
we added the `deeply bound' 1$s$ state of $^{205}$Pb \cite{GGG02} or,
alternatively, the `normal' 1$s$ states in the neighbouring $N>Z$ nuclei
$^{18}$O and $^{22}$Ne, in order to determine the isovector amplitude $b_1$.
With fits to such greatly reduced data sets one must assume
a fixed $p$-wave potential, which we took from Ref. \cite{Fri02a}.
The results of best-fit values for $b_1$ using potential C within these
partial data sets are shown in Table \ref{tab:fits}, together with similar
results using potential C within global data sets: `global 1' from Ref.
\cite{Fri02a} and `global 2' which extends it by including also
$^{12,13}$C and $^{14}$N.
It is clear that, with the fairly small errors on $b_1$ in the global fits,
the best-fit value of $b_1$ is enhanced over 20\% with respect to the
free-space value of $b_1 = -0.0885^{+0.0010}_{-0.0021}~m_\pi ^{-1}$
\cite{SBG01}, whereas
the larger errors associated with using restricted data sets can hardly
qualify for making such a statement. We would like to emphasize that,
with the realistic uncertainties which are typical of the
restricted data sets, one cannot argue for a solid evidence of in-medium
enhancement of $b_1$.
It is interesting to note, for the restricted fits, that smaller errors are
obtained using the two neighbouring $^{18}$O and $^{22}$Ne than
using the deeply bound $^{205}$Pb. This suggests that the `deeply bound'
pionic $1s$ states do not carry new information over that already contained
in the `normal' pionic $1s$ states, as demonstrated below.

\begin{table}
\caption{Values of $b_1$ from fits to several data sets using potential C.}
\label{tab:fits}
\begin{tabular}{lcccc}
\hline
data & light $N=Z$ & light $N=Z$ & global 1 & global 2 \\
 &$+^{205}$Pb &+ light $N>Z$ & $^{16}$O to $^{238}$U &$^{12}$C to $^{238}$U \\
\hline
points       & 14 & 16 & 106 & 114 \\
  $\chi ^2 /F$  & 3.0 & 2.9 & 1.9 & 2.1 \\
$b_1 (m_\pi ^{-1})$ &$-$0.113$\pm$0.025 &$-$0.096$\pm$0.014&
$-$0.113$\pm$0.004 &$-$0.112$\pm$0.004 \\
\hline
\end{tabular}\\[2pt]
\end{table}

\subsection{The role of deeply bound pionic atom states}

Finally we remark on the role played by the deeply bound 1$s$ and 2$p$
states of pionic atoms in providing information on the pion-nucleus
interaction. Global fits have shown that the 1$s$ and 2$p$
`deeply bound' states in $^{205}$Pb follow the general trend observed
for more than 100 `normal' states and that the agreement between calculation
and experiment for them was actually better than the average
(Fig. 3 of Ref. \cite{Fri02a}).
Similar conclusions were obtained earlier for $^{207}$Pb \cite{FGa98}.

\begin{figure}[bht]
\epsfig{file=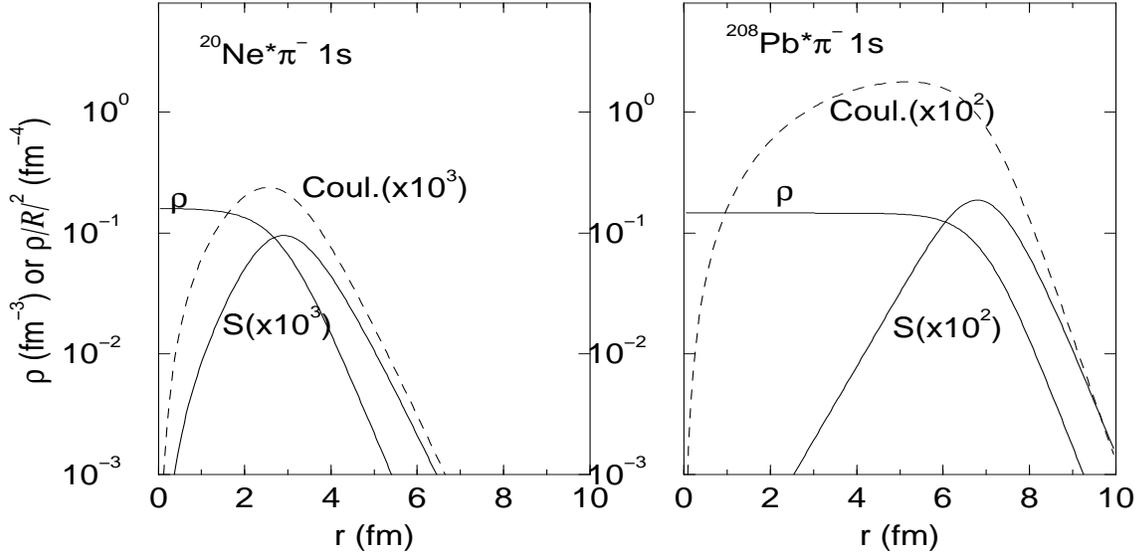,height=75mm,width=150mm}
\caption{Radial densities for pionic $1s$ states  
multiplied by the nuclear density:
dashed for (finite size) Coulomb potential, solid curves for Coulomb plus
strong interaction.}
\label{fig:over}
\end{figure}

Figure \ref{fig:over} shows that this conclusion follows naturally from
the properties of the atomic wavefunctions and their relation to the nuclear
density. The figure displays absolute values squared of the radial
wavefunctions multiplied by the nuclear density for a normal 1$s$ state
(in $^{20}$Ne) and for a deeply bound 1$s$ state (in $^{208}$Pb).
It is seen that the Coulomb wavefunction would have
indeed penetrated deeper into the heavy Pb nucleus, but due to the strong
interaction it is repelled such that its overlap with
the nucleus is sufficiently small to make the width of the state
relatively narrow and thus making the state observable. In fact, the
`deeply bound' 1$s$ wavefunction does not overlap with inner regions of the
nucleus more so than a `normal' 1$s$ wavefunction does.
It is thus concluded that deeply bound states do not play any special role
in the determination of pionic atom potentials.
This is not surprising since the same mechanism which causes the deeply bound
states to be narrow, namely, the strong repulsion \cite{FSo85,TYa88} of
the wavefunction out of the nucleus, also masks the nuclear interior such
that the penetration of the deeply bound pionic atom wavefunction is not
dramatically enhanced compared to the `normal' states.


\begin{thebibliography}{9}

\bibitem{BFG97}For a recent review see  C.J. Batty, E. Friedman, A. Gal,
Phys. Rep. 287 (1997) 385.

\bibitem{Wei01}W. Weise, Nucl. Phys. A 690 (2001) 98c.

\bibitem{KKW02}E.E. Kolomeitsev, N. Kaiser and W. Weise, nucl-th/0207090.

\bibitem{CEO02}G. Chanfray, M. Ericson and M. Oertel, nucl-th/0211035.

\bibitem{Fri02}E. Friedman, Phys. Lett. B 524 (2002) 87.

\bibitem{Fri02a}E. Friedman, Nucl. Phys. A 710 (2002) 117.

\bibitem{SBG01}H.-Ch. Schr\"oder et al., Eur. Phys. J. C 21 (2001) 473.

\bibitem{KYa01}P. Kienle and T. Yamazaki, Phys. Lett. B 514 (2001) 1.

\bibitem{GGG02a}H. Geissel et al., Phys. Lett. B 549 (2002) 64; 
see also T. Yamazaki and S. Hirenzaki, nucl-th/0210040.

\bibitem{GGG02}H. Geissel et al., Phys. Rev. Lett. 88 (2002) 122301.

\bibitem{FGa98}E. Friedman and A. Gal, Phys. Lett. B 432 (1998) 235.

\bibitem{FSo85}E. Friedman and G. Soff,
J. Phys. G: Nucl. Phys. 11 (1985) L37.

\bibitem{TYa88}H. Toki and T. Yamazaki, Phys. Lett. B 213 (1988) 129.

\end{thebibliography}
\end{document}